\documentclass[a4paper]{jpconf}
\usepackage{graphicx}
\usepackage{enumerate}
\usepackage{amssymb, amsmath, amsfonts, amsthm}
\usepackage[utf8]{inputenc} 
\usepackage{float}
\usepackage{graphicx}
\usepackage{flushend}
\usepackage{multicol}

\usepackage{multirow}
\usepackage{slashed}
\usepackage{mathrsfs}
\newcommand{\la}{\langle}
\newcommand{\ra}{\rangle}

\newcommand{\g}{\gamma}
\newcommand{\ba}{\begin{align}}
\newcommand{\ea}{\end{align}}
\newcommand{\be}{\begin{equation}}
\newcommand{\ee}{\end{equation}}

\begin{document}

\title{An Introduction to the Massive  Helicity Formalism with applications to the Electroweak SM}

\author{J. Lorenzo D\'iaz-Cruz$^{1}$, Bryan O. Larios$^{2}$ and O. Meza-Aldama$^{3}$}

\address{$^{1}$ Facultad de Ciencias F\'isico - Matem\'aticas, BUAP \protect\\Apdo. Postal 1364, C.P. 72000, Puebla, Pue. M\'exico}

\ead{$^1$ldiaz@ifuap.buap.mx, $^2$bryanlarios@gmail.com, $^3$oscar.meza@email.ucr.edu}



\begin{abstract}
The power of the helicity formalism has been appreciated recently from its application to the massless case, where plenty of formal aspects of multi-legs amplitudes have been derived. However, in order to extend the formalism to the realistic cases, such as the electroweak Standard Model and QCD with massive quarks, some extra inputs are needed. We discuss first 
the formalism needed to  evaluate amplitudes expressed with the modern notation  for massive fermions and vectors based on a thorough   treatment of Proca vector fields. Then, we present some examples of elementary processes where it is shown how the formalism leads to tremendous simplifications, these include 2-body decays $Z\to ff$, $h\to ff$ and $h\to W^{-}W^{+}$ as well as the 3-body decay 
$h\to V f'\bar{f}$.
\end{abstract}
\section{Introduction}
The spinor helicity formalism (SHF) is a tool for calculating scattering amplitudes much more efficiently than the traditional approach. Its versatility is based on the fact that all objects appearing in the Feynman rules for quantum field theories such as QED, QCD, EW and SUSY can be written in terms of two-component Weyl spinors (and Pauli sigma matrices). Dotted and undotted Weyl spinors are used, and in some sense they could be considered more fundamental than Minkowski four-vectors since they form irreducible representations of the Lorentz group,  unlike Dirac spinors which are a ``mixture'' (i.e. a direct sum) of two different representations. 
\indent When dealing with a massless fermion, the solutions of the momentum-space Dirac equation only has two non-vanishing components, 
in that case the distinction between a 4-component Dirac spinor and a 2-component Weyl spinor essentially disappears. 
These Weyl spinors, which are helicity eigenstates in the massless case, can then be used to rewrite the Feynman rules for external legs (both fermionic and bosonic), vertices and propagators. 
The expressions obtained for the helicity amplitudes are in general very simple, and they can be squared directly, without any need of spinor completeness relations and Casimir tricks. 
This is one of the most marvelous advantages of both massless and massive SHF. \newline
\indent The SHF  usually concentrates  on  the massless case (see references \cite{srednicki}-\cite{peskin2}),  while the massive formalism has been less studied in the current literature. The only notable exception is Ref. \cite{dittmaier}, although it is written in a somewhat old-fashioned notation. Other papers, such as \cite{badger1}-\cite{dixon}, study scattering amplitudes of several theories using mostly BCFW recursion (Britto-Cachazo-Feng-Witten) \cite{bcfw}, creating a  gap between the well-understood massless SHF and the  work done in the massive SHF even for ``simpler'' theories such as QED. \newline
\indent 
In the massive case, the simplifications are not that large, the method is in general a little bit more tricky than the massless case \cite{us}. 
Finally, we must say that one of the main applications of the modern amplitude calculation techniques is to QCD gluon processes. Using recursion relations such as BCFW, one can calculate multi-gluon scattering amplitudes which would be impossible (in practice) to obtain using ``traditional'' methods. Of course, gluons are massless, so one does not need to extend the SHF to include massive particles in this case. \newline
\indent Computational tools have been developed to deal with both massless and massive spinor helicity amplitudes (see Refs. \cite{sam} and \cite{spinorsextras}), which of course comes very handy when dealing with complicated processes. However, one can notice that a self-contained theory of massive helicity amplitudes and detailed applications  to simple processes of realistic theories is  lacking, which is one of the motivations of this paper.

In this work we show some practical examples where we have applied the massive SHF to several processes (at tree level) in the Electroweak SM. 
This paper is structured as follows:  Section \ref{section2} contains the solutions of the Dirac equation (massive case), as well the solutions of the Maxwell and Proca equations in order to derive the Feynman rules for external fermionic and bosonic lines, then we use the so-called light cone decomposition (LCD) \cite{spinorsextras,us,kosower} to express massive fermions in terms of massless ones.  In Section \ref{section3} we present some calculations in Electroweak SM. Finally in Section \ref{conclussions} we present our conclusions and final comments of this work. 

\section{Helicity Method for  QED}\label{section2}
Consider a theory with a massive Dirac fermion, which is described by Dirac equation. Writing the momentum space Dirac spinor $u$ in terms of two Weyl spinors the equation of motion takes the form
\begin{equation}\label{pspace_dirac_eq}
	(\slashed{p}+m)u(p)=
	\left(\begin{array}{cc}
		m & p_\mu\sigma^\mu_{a\dot{a}} \\
		p_\mu\bar{\sigma}^{\mu\dot{a}a} & m
	\end{array}\right)
	\left(\begin{array}{c}
		\chi_a(p) \\ \xi^{\dot{a}}(p)
	\end{array}\right)
	= 0.
\end{equation}
Through all the paper we shall use the following spacetime metric $g^{\mu\nu}=\text{diag}(-1,1,1,1)$, also we have used the base where the Dirac matrices take the following form
\begin{equation}\label{diracgamma}
\g^{\mu}=\left(\begin{array}{c c }
		0 & \sigma^{\mu} \\ 
		\bar{\sigma}^{\mu} & 0
	\end{array}\right),
\end{equation}
with $\sigma^{\mu}=(1,\vec{\sigma})$ and $\bar{\sigma}^{\mu}=(1,-\vec{\sigma})$.\newline

Weyl spinor ($\xi^{\dot{a}}$, $\chi_a$) could be expressed in terms of brackets (angle and square) spinors, they read as  $\chi=|r], \, \xi=\frac{m}{\langle rq\rangle}|q\rangle$ 
where  $|r]$ and $|q\ra$ are  2-component Weyl spinors linked to the light-like momenta $r^\mu$ and $q^{\mu}$, with the spinor indices being omitted.
 They arise when one uses LCD to express a massive momentum ($p^\mu$) in term of  2 massless momenta ($p^\mu=r^\mu+\frac{p^2q^\mu}{2r\cdot q}$), 
this approach will be used throughout the paper to express massive spinors in terms of massless ones. 

Equation  (\ref{pspace_dirac_eq}) has 4 solutions,  (see Ref. \cite{us} for a complete derivation), these are: 
\begin{equation}\label{equ:14}
	u_-(p)=|r]+\frac{m}{\la rq\ra}|q\ra
	, \quad
	u_+(p)=\frac{m}{[rq]}|q]+|r\ra ,
\end{equation}
\begin{equation}
	v_+(p)=|r]-\frac{m}{\la rq\ra}|q\ra
	, \quad
	v_-(p)=-\frac{m}{[rq]}|q]+|r\ra ,
\end{equation}
with the conjugate solutions given by
\begin{equation}
	\bar{u}_-(p)=\frac{m}{[qr]}[q|+\la r|, \quad \bar{u}_+(p)=[r|+\frac{m}{\la qr\ra}\la q|,
\end{equation}
\begin{equation}
	\bar{v}_+(p)=-\frac{m}{[qr]}[q|+\la r|, \quad \bar{v}_-(p)=[r|-\frac{m}{\la qr\ra}\la q|. \label{equ:17}
\end{equation}

Spinor products satisfy $[rq]\equiv-[qr]$, similarly $\langle rq\rangle=-\langle qr\rangle$.
For real momenta we have $\langle pk\rangle\equiv[kp]^\ast
$. When $m=0$, one can verify that $\bar{u}_\pm=\bar{v}_\mp$.
Equations (\ref{equ:14})-(\ref{equ:17}) will allow us to compute scattering amplitudes in a more practical way than  the traditional approach.
For massless spin-1 particles, one can write the transversal polarization vectors as follows $\epsilon_+^\mu(p) = - \frac{\la q|\gamma^\mu|p]}{\sqrt{2}\la qp \ra}$ and $\epsilon_-^\mu(p) = - \frac{\la p|\gamma^\mu|q]}{\sqrt{2}[qp]}$, see Ref.\cite{srednicki,elvang}, where $p$ is the spin-1 momentum, and $q$ is a reference momentum defined such that $q^2=0$.
Note that $(\epsilon_\pm^\mu)^\ast (p) = - \epsilon_\mp^\mu (p)$. The slashed versions of these polarization vectors will also be useful to compute scattering amplitudes in the examples presented later, these are $\slashed{\epsilon}_+(p) = \frac{\sqrt{2}}{\la qp \ra} \left( |p] \la q| + |q \ra [p| \right)$ and $\slashed{\epsilon}_-(p) = \frac{\sqrt{2}}{[qp]} \left( |p\ra [q| + |q] \la p| \right) $.

\indent Finally, for massive spin-1 bosons (see Ref.~\cite{spinorsextras, us}) making the identification $\epsilon_{1,2,3}^\mu(p) \equiv \epsilon_{-,+,0}^\mu$(p), we can write $\epsilon_+^\mu(p) = \frac{\la r|\gamma^\mu|q]}{\sqrt{2}[ rq]}$, $\epsilon_-^\mu(p) = \frac{\la q|\gamma^\mu|r]}{\sqrt{2}\la rq\ra} $ and finally the longitudinal mode $ \epsilon_0^\mu(p)  = \frac{1}{m}r^\mu + \frac{m}{2 p \cdot q} q^\mu$,  where now $p^2=-m^2$, the momenta $r$ and $q$ are associated with the momentum $p$ by LCD.



\section{Elementary Processes in the Standard Model}\label{section3}
In this section we will use the SHF to compute several elementary processes in the Standard Model of particle physics. Considering massive particles but using LCD, we shall express the massive 4-component Dirac spinor in terms of the massless 2-component Weyl spinors. This will allow us to exploit all the  available identities from the massless SHF.  
\subsection{2-body Higgs decay $h(p_1)\to W^{+}(p_2)W^-(p_3)$}
The helicity amplitude (HA) for the process $h\to W^{+}W^-$ reads as follows
\begin{align}\label{eq:2001}
\mathcal{M}_{\lambda_{2}\lambda_{3}}(p_1,p_2,p_3)&=\frac{2M_{W}^2g_{\mu\nu}}{v}\epsilon_{\lambda_2}^{\mu}(p_2)\epsilon^\nu_{\lambda_3}(p_3),
\end{align}
where $p_1^2=-M_h^2$, $p_2^2=p_3^2=-M_W^2$ and the $\lambda$'s represent the helicity of the particles. We shall use simultaneous light cone decomposition (SLCD) to express the  massive momenta $p_2$ and $p_3$ in terms of massless momenta ($r_2,\,q_2,\,r_3,\,q_3$). The massive momenta take the form; $p_2=r_2-\frac{M_W^2}{2r_2\cdot q_2}q_2\label{equ:01},$ and $p_3=q_2-\frac{M_W^2}{2r_2\cdot q_2}r_2\label{equ:02}$, here SLCD implies that $r_3=q_2$ and $q_3=r_2$. 
 For this process there are $3^2$ HA's, $\mathcal{M}_{++},\,\mathcal{M}_{+-},\,\mathcal{M}_{-+}$, $\mathcal{M}_{--},\,\mathcal{M}_{0+},\,\mathcal{M}_{0-},\mathcal{M}_{-0},\,\mathcal{M}_{+0}$ and $\mathcal{M}_{00}$. Using SLCD to fix the massless momenta will be crucial to reduce the number of  HA's, in this  case HA's: $\mathcal{M}_{+-},\,\mathcal{M}_{-+},\,\mathcal{M}_{0-},\,\mathcal{M}_{-0},\,\mathcal{M}_{+0},\,\mathcal{M}_{0+}$ vanish. The nonzero HA's are shown in Table \ref{table1}.
\begin{table}[H]
\begin{center}
  \begin{tabular}{  || c |  c || }
    \hline \hline
    $\lambda_2\lambda_3$ &$\mathcal{M}_{\lambda_2\lambda_{3}}$ \\ \hline
   \hline\hline
     $--$ & $\frac{2M_W^2}{v}\frac{\la q_3q_2\ra}{[q_3q_2]}$\\ \hline
     $++$ & $\frac{2M_W^2}{v}\frac{[q_3q_2]}{\la q_3q_2\ra}$ \\ \hline
          $00$ & $-\frac{(2M_W^2)(s_{q_2q_3}^2+M_W^4)}{vM_W^2s_{q_2q_3}}$ \\ \hline
    \hline
  \end{tabular}
  \caption{Helicity Amplitudes for the 2-body Higgs decay  $h\to W^{+}W^-$.}
  \label{table1}
 \end{center}
\end{table}
In the expressions of Table \ref{table1} we have used the following spinor relations
\begin{align}
	[q_2|\gamma^\mu|q_3\rangle &=\langle q_3|\gamma^\mu|q_2] , \label{sq_br_1} \\
	[q_2|\gamma^\mu|q_3\rangle^\ast &=[q_3|\gamma^\mu|q_2\rangle  \label{sq_br_2},
\end{align}
as well as Fierz identity:
\begin{equation}\label{equ:fierz}
	\langle q_2|\gamma^\mu|r_2]\langle r_3|\gamma_\mu|q_3]=2\langle q_2r_3\rangle[r_2q_3] .
\end{equation}
The averaged and squared amplitude for the 2-body Higgs decay ($h\to W^{+}W^-$) is: 
\begin{align}
\left\la|\mathcal{M}|^2\right\ra&=2|\mathcal{M}_{++}|^2+|\mathcal{M}_{00}|^2,\label{equ:3002}\\
&=\left(\frac{2M_W^2}{v}\right)^2\left(2+\left(\frac{1}{2M_W^2s_{q_2q_3}}\right)^2(s_{q_2q_3}^2+M_W^4)^2\right),\\
&=\frac{M_h^4}{v^2}\left(1-4x^2+12x^4\right),\label{eq:2002}
\end{align}
with $s_{q_2q_3}=-(q_2+q_3)^2=-2q_2\cdot q_3$ and $x=\frac{M_W}{M_h}$, the momenta $q_2$ and $q_3$ are defined as follows \cite{spinorsextras,us}
\begin{equation}\label{equ:masslessmomentum}
q_2=\frac{\big(\text{sgn}(p_2\cdot p_3)\sqrt{\Delta}+p_2\cdot p_3\big)p_3-p_3^2\,p_2}{2\,\text{sgn}(p_2\cdot p_3)\sqrt{\Delta}}\hspace{.15cm}\text{and}\hspace{.15cm}q_3=\frac{\big(\text{sgn}(p_2\cdot p_3)\sqrt{\Delta}+p_2\cdot p_3\big)p_2-p_2^2\,p_3}{2\,\text{sgn}(p_2\cdot p_3)\sqrt{\Delta}},
\end{equation}
with $\Delta=(p_2\cdot p_3)^2-p_2^2\,p_3^2$. 
Furthermore we have used in Eq.~(\ref{equ:3002}) that $|\mathcal{M}_{--}|^2=|\mathcal{M}_{++}|^2$. From Eq. (\ref{eq:2002}) we find  the decay width $\Gamma$ for the process $h\to W^{+}W^-$:\begin{align}\label{equ:03}
\Gamma(h\to W^{+}W^-)=\frac{\lambda(1,x,x)^{1/2}}{16\pi M_h}\left\la|\mathcal{M}|^2\right\ra=\frac{\alpha_WM_h}{16x^2}\left(1-4x^2+12x^4\right)\sqrt{1-4x^2},
\end{align}
where  $\alpha_W=\frac{M_W^2}{v^2\pi}$ and the term $\sqrt{1-4x^2}$ is the $W$ velocity in the Higgs rest reference frame.
\subsection{2-body Higgs decay $h(p_1)\to f(p_2)\bar{f}(p_3)$}
The HA for the process $h\to f\bar{f}$ is the following:
\begin{align}\label{eq:2003}
\mathcal{M}_{\lambda_2\lambda_3}(p_1,p_2,p_3)&=\frac{1}{v}\bar{u}_{\lambda_2}(p_2)v_{\lambda_3}(p_3),
\end{align} 

where $p_1^2=-M_h^2$, $p_2^2=p_3^2=-m_f^2$. 
There are $2^2$ HA's, $\mathcal{M}_{+-},\,\mathcal{M}_{-+},\,\mathcal{M}_{--}$ and $\mathcal{M}_{++}$, but using SLCD for momenta $p_2$ and $p_3$,  $\mathcal{M}_{-+}$ and $\mathcal{M}_{+-}$ vanish. The nonzero HA's are shown in Table \ref{table2}.
\begin{table}[H]
\begin{center}
  \begin{tabular}{  || c |  c || }
    \hline \hline
    $\lambda_2\lambda_3$ &$\mathcal{M}_{\lambda_2\lambda_{3}}$ \\ \hline
   \hline\hline
     $--$ & $\frac{m_f}{v[q_2q_3]}(s_{q_2q_3}-m_f^2)$\\ \hline
     $++$ & $\frac{m_f}{v\la q_2q_3\ra}(s_{q_2q_3}-m_f^2)$ \\ \hline
    \hline
  \end{tabular}
  \caption{Helicity Amplitudes for the Higgs decay  $h\to f\bar{f}$.}
  \label{table2}
 \end{center}
\end{table}
The averaged squared amplitude is then:
\begin{align}
\left\la|\mathcal{M}|^2\right\ra&=2|\mathcal{M}_{--}|^2=2|\mathcal{M}_{++}|^2
=\frac{2m_f^2}{v^2s_{q_2q_3}}(s_{q_2q_3}-m_f^2)^2
=\frac{y^2}{v^2}(1-4y^2),\label{eq:2004}
\end{align}
with $y=\frac{m_f}{M_h}$. 
Then the decay width $\Gamma$ goes as follows
\begin{align}
\Gamma(h\to f\bar{f})=\frac{\alpha_WM_h y^2}{8}(1-4y^2)^{3/2}.
\end{align}
\subsection{2-body $Z$ boson decay $Z(p_1) \rightarrow f(p_2) \bar{f}(p_3)$}
The HA for the process $Z \rightarrow f \bar{f}$ is given as follows
\begin{equation}
	{\cal M}_{\lambda_1\lambda_2\lambda_3} = \frac{1}{2} g_Z \epsilon_{\mu\lambda_1} (p_1) \bar{u}_{\lambda_2} (p_2) \gamma^\mu (v_f - a_f \gamma^5) v_{\lambda_3}(p_3) .
\end{equation}
We shall assume that $m_f \ll M_Z \equiv M$, that could not true   when the fermion is the quark top, while for the other cases in practice it will be equivalent to consider massless fermions  ($p_2^2=p_3^2=0$). Then the amplitude will vanish unless $f$ and $\bar{f}$ have opposite helicities. We choose the arbitrary reference momentum $q_1 = p_2$, then by momentum conservation ($\sum_{i=1}^n[qi]\langle ik\rangle=0$)
\begin{equation}\label{equ:08}
	\la p_3 r_1 \ra [r_1p_2] = 0,
\end{equation}
Using  Eq.(\ref{equ:08}), the independent HA's are shown in Table \ref{table3}.
\begin{table}[H]
\begin{center}
  \begin{tabular}{  || c |  c || }
    \hline \hline
    $\lambda_1\lambda_2\lambda_3$ &$\mathcal{M}_{\lambda_1\lambda_2\lambda_{3}}$ \\ \hline
   \hline\hline
     $++-$ & $\frac{1}{2} g_Z \frac{\la p_2| \gamma_\mu | r_1]}{\sqrt{2} \la r_1p_2 \ra} (v_f - a_f) [p_2|\gamma^\mu|p_3\ra = \frac{g_Z (v_f - a_f)}{\sqrt{2}} \frac{\la p_2p_3 \ra [r_1p_2]}{\la r_1p_2 \ra} $\\ \hline
     $0+-$ & $\frac{1}{2} g_Z \left( \frac{1}{M}{r_1}_\mu + \frac{M}{2p_{12}}{p_2}_\mu \right) (v_f - a_f) [p_2| \gamma^\mu |p_3\ra = \frac{g_Z (v_f - a_f)}{\sqrt{2} M} \la r_1p_3 \ra [r_1p_2] = 0 $\\ \hline 
     $-+-$ & $\frac{1}{2} g_Z \frac{\la r_1 |\gamma_\mu| p_2]}{\sqrt{2} [p_2r_1]} (v_f - a_f) [p_2 |\gamma^\mu |p_3\ra = 0 $
     \\ \hline
    \hline
  \end{tabular}
  \caption{Helicity Amplitudes for the Higgs decay  $h\to f\bar{f}$.}
  \label{table3}
 \end{center}
\end{table}
The rest of the HA's are obtained by complex conjugation. Besides, because of the  $\gamma^5$ matrix,  the sign of the coefficient $a_f$ will change. This is because $\gamma^5 v_-(p) = v_-(p)$ but $\gamma^5 v_+ (p) = - v_+ (p)$. Then we have
\begin{equation}
	{\cal M}_{--+} = \frac{g_Z (v_f + a_f)}{\sqrt{2}} \frac{[p_2p_3] \la r_1p_2 \ra}{[r_1p_2]} .
\end{equation}
\indent Taking the square moduli of the nonzero amplitudes we obtain
\begin{gather}
	\left| {\cal M}_{++-} \right|^2 = - g_Z^2 \left( |v_f|^2 + |a_f|^2 - 2 \text{Re} (v_f a_f^\ast) \right) p_{23} , \\
	\left| {\cal M}_{--+} \right|^2 = - g_Z^2 \left( |v_f|^2 + |a_f|^2 + 2 \text{Re} (v_f a_f^\ast) \right) p_{23} .
\end{gather}
From momentum conservation $p_1=p_2+p_3$, then we obtain: $p_2\cdot p_3=p_{23}=-\frac{1}{2}M^2$. Finally
\begin{equation}
	\left\langle \left| {\cal M} \right|^2 \right\rangle = \frac{1}{3} \left( \left| {\cal M}_{++-} \right|^2 + \left| {\cal M}_{--+} \right|^2 \right) = \frac{g_Z^2 M^2}{3} \left( |v_f|^2 + |a_f|^2 \right) .
\end{equation}
The decay width for this channel is then:
\begin{equation}
	\Gamma \left( Z \rightarrow f\bar{f} \right) = \frac{g_Z^2 M}{48 \pi} \left( |v_f|^2 + |a_f|^2 \right) .
\end{equation}


\subsection{3-body  Muon Decay  $\mu(p_1)\to\bar{\nu}_{e^-}(p_2)\,\nu_{\mu}(p_3)\,e^-(p_4)$}
The HA for the process $\mu\to\bar{\nu}_{e^-}\,\nu_{\mu} e^-$ is as follows
\begin{align}\label{eq:201}
\mathcal{M}_{\lambda_1\lambda_2\lambda_3\lambda_4}&=\left(\frac{g_W}{\sqrt{8}M_W}\right)^2\left[\bar{u}_{\lambda_3}(p_3)\gamma^{\mu}(1-\gamma_5)u_{\lambda_1}(p_1)\right]\left[\bar{u}_{\lambda_4}(p_4)\gamma_{\mu}(1-\gamma_5)v_{\lambda_2}(p_2)\right],\\
&=\left(\frac{g_W}{\sqrt{8}M_W}\right)^2\mathcal{A}^{\mu}_{\lambda_3\lambda_1}\mathcal{B}_{\mu\lambda_4\lambda_2},
\end{align}
where $p_1^2=-m_{\mu}^2$, $p_4^2=-m_e^2$, $p_2^2=0$ and $p_3^2=0$. We have defined $\mathcal{A}^{\mu}_{\lambda_3\lambda_1}\,\text{and}\,\mathcal{B}_{\mu\lambda_4\lambda_2}$ as follows
\begin{align}\label{eq:202}
\mathcal{A}^{\mu}_{\lambda_3\lambda_1}&=2\bar{u}_{\lambda_3}(p_3)\gamma^{\mu}\hat{P}_Lu_{\lambda_1}(p_1),\\
\mathcal{B}_{\mu\lambda_4\lambda_2}&=2\bar{u}_{\lambda_4}(p_4)\gamma_{\mu}\hat{P}_Lv_{\lambda_2}(p_2),\label{eq:2031}
\end{align}
From equations (\ref{eq:202}) and (\ref{eq:2031})  one obtains  the following HA's; $\mathcal{A}^{\mu--}=0,\,\mathcal{A}^{\mu-+}=0,\,\mathcal{A}^{\mu+-}=\frac{2m_{\mu}}{\la r_1q_1\ra}[p_3|\g^{\mu}|q_1\ra,\, \mathcal{A}^{\mu++}=2[p_3|\gamma^{\mu}|r_1\ra,\,\text{and}\,\mathcal{B}_{\mu}^{++}=0,\,\mathcal{B}_{\mu}^{--}=0,\,\mathcal{B}_{\mu}^{-+}=\frac{2m_e}{[q_4r_4]}[q_4|\g_{\mu}|p_2\ra,\,\mathcal{B}_{\mu}^{+-}=2[r_4|\g_{\mu}|p_2\ra$, in  Table \ref{table4} we show the nonzero products for the partial helicity amplitudes $\mathcal{A}^{\mu\lambda_3\lambda_1}$ and $\mathcal{B}_{\mu}^{\lambda_4\lambda_2}$.
\begin{table}[H]
\begin{center}
  \begin{tabular}{  || c | c | c | c | c || }
    \hline \hline
    $\lambda_1\lambda_2\lambda_3\lambda_4$ &$\mathcal{A}^{\mu\lambda_3\lambda_1}$ & $\mathcal{B}_{\mu}^{\lambda_4\lambda_2}$ & $\mathcal{A}^{\mu\lambda_3\lambda_1}\mathcal{B}_{\mu}^{\lambda_4\lambda_2}$ &  $\mathcal{M}\,\,$ $(p_2=q_1,\,p_3=q_4)$\\ \hline
   \hline\hline
     $+-+-$& $2[p_3|\g^{\mu}|r_1\ra$ & $2[r_4|\g_{\mu}|p_2\ra $ & $4\la p_2r_1\ra \la p_3r_4\ra$ & $\left(\frac{g_W}{\sqrt{2}m_{\mu}}\right)^2\la p_2r_1\ra[p_3r_4]$\\ \hline 
    $+++-$ & $2[p_3|\g^{\mu}|r_1\ra$ & $\frac{2m_e}{[q_4r_4]}[q_4|\g_{\mu}|p_2\ra$ &$4\frac{m_e}{[q_4r_4]}\la p_2r_1\ra\la p_3q_4\ra$ & 0\\ \hline
   $--++ $ & $ \frac{2m_{\mu}}{\la r_1q_1\ra}[p_3|\g^{\mu}|q_1\ra$ & $ 2[r_4|\g_{\mu}|p_2\ra$ & $4\frac{m_{\mu}}{\la r_1q_1\ra}\la p_2q_1\ra[p_3 r_4]$ & 0\\ \hline
   $-++-$ & $\frac{2m_{\mu}}{\la r_1q_1\ra}[p_3|\g^{\mu}|q_1\ra$ & $\frac{2m_e}{[q_4r_4]}[q_4|\g_{\mu}|p_2\ra$ & $4\frac{m_{\mu}m_e}{\la r_1q_1\ra[q_4r_4]}\la p_2q_1\ra[p_3q_4]$ & 0\\
    \hline\hline
  \end{tabular}
  \caption{Helicity Amplitudes for Muon decay ($\mu\to e^-\,\bar{\nu}_{e^-}\,\nu_{\mu}$) with the momentum assignment $r_2=q_1$ and $r_3=q_4$.}
  \label{table4}
 \end{center}
\end{table}
We have used in the fourth column of the Table \ref{table4} the reflection property Eq.~(\ref{sq_br_1}) and Fierz identity Eq.~(\ref{equ:fierz}), while in the last column it was  chosen $r_2=q_1$ and $r_2=q_4$, this reduces all the helicity amplitudes to one ($\mathcal{M^{+-+-}}$). The squared and averaged amplitude for the muon decay is:\begin{align}\label{eq:203}
\la|\mathcal{M}^{+-+-}|^2\ra=\frac{1}{2}|\mathcal{M^{+-+-}}|^2=2\left(\frac{g_W}{M_W}\right)^4(p_1\cdot p_2)(p_3\cdot p_4)
\end{align} 
From this result we can arrive to the decay width, which agrees with result of textbooks.
\subsection{3-body Higgs decay $h(p_1) \rightarrow Z(p_2) f(p_3) \bar{f}(p_4)$}
The HA for this process is the following
\begin{equation}\label{equ:12}
	{\cal M}_{\lambda_2\lambda_3\lambda_4} = \frac{M_Z^2}{v} \frac{e}{c_W s_W} \epsilon^\mu_{\lambda_2} (p_2) \left( g_{\mu\nu} + \frac{k_\mu k_\nu}{M_Z^2} \right) \frac{1}{k^2 + M_Z^2} \bar{u}_{\lambda_3} (p_3) \gamma^\nu (v_f - a_f \gamma^5) v_{\lambda_4}(p_4) .
\end{equation}
We will consider just the case when the fermions are leptons ($e,\mu, \tau,\nu_{e},\nu_{\mu},\nu_{\tau}$), with this approximation the leptons are consider as massless, so they must have opposite helicities. Then, remembering that $k = p_3 + p_4$, we have
\begin{gather}\label{equ:10}
	k_\nu \bar{u}_+ (p_3) \gamma^\nu (v_f - a_f \gamma^5) v_-(p_4) = (v_f - a_f) [p_3| (\slashed{p}_3 + \slashed{p}_4) |p_4\ra = 0 , \\
	k_\nu \bar{u}_- (p_3) \gamma^\nu (v_f - a_f \gamma^5) v_+(p_4) = (v_f + a_f) \la p_3| (\slashed{p}_3 + \slashed{p}_4) |p_4] = 0 ;\label{equ:11}
\end{gather}
it is very hard to find this kind of simplification (Equations (\ref{equ:10}) and (\ref{equ:11})) with the traditional approach of computing scattering amplitudes, but now the amplitude (\ref{equ:12}) becomes simply 
\begin{equation}
	{\cal M}_{\lambda_2\lambda_3\lambda_4} = {\cal C} \bar{u} (p_3) \slashed{\epsilon} (p_2) (v_f - a_f \gamma^5) v (p_4) ,
\end{equation}
where we have defined ${\cal C} \equiv \frac{M_Z^2}{v} \frac{e}{c_W s_W} \frac{1}{k^2 + M_Z^2}$. Choosing $q_2 = p_3$, we get

\begin{align}
	\slashed{\epsilon}_+ (p_2) &= \frac{\sqrt{2}}{\la p_3r_2 \ra} (|r_2] \la p_3| + |p_3\ra [r_2|) , \\
	\slashed{\epsilon}_- (p_2) &= \frac{\sqrt{2}}{[ r_2p_3 ]} (|p_3] \la r_2| + |r_2 \ra [p_3|) , \\
	\slashed{\epsilon}_0 (p_2) &= \frac{1}{M_Z} \slashed{r}_2 + \frac{M_Z}{2 p_{23}} \slashed{p}_3 .
\end{align}
The nonzero HA's are show below in the Table \ref{table5}.
\begin{table}[H]
\begin{center}
  \begin{tabular}{  || c |  c || }
    \hline \hline
    $\lambda_2\lambda_3\lambda_4$ &$\mathcal{M}_{\lambda_2\lambda_3\lambda_{4}}$ \\ \hline
   \hline\hline
     $++-$ & ${\cal C} [ p_3 | \frac{\sqrt{2}}{\la p_3 r_2 \ra} | r_2 ] \la p_3 | (v_f - a_f \gamma^5) | p_4 \ra = \sqrt{2} {\cal C} (v_f - a_f) \frac{[ p_3 r_2 ] \la p_3 p_4 \ra}{\la p_3 r_2 \ra}  $\\ \hline
     $--+$ & ${\cal C} \la p_3 | \frac{\sqrt{2}}{[ r_2 p_3 ]} | r_2 \ra [ p_3 | (v_f - a_f \gamma^5) | p_4 ] = \sqrt{2} {\cal C} (v_f + a_f) \frac{\la p_3 r_2 \ra [ p_4 p_3 ]}{[ p_3 r_2 ]}$\\ \hline 
     $0+-$ & ${\cal C} [ p_3 | \left( \frac{1}{M_Z^2} \slashed{r}_2 + \frac{M_Z}{2 p_{23}} \slashed{p}_3 \right) (v_f - a_f \gamma^5 ) | p_4 \ra = {\cal C} \frac{v_f - a_f}{M_Z} [ r_2 p_3 ] \la r_2 p_4 \ra$\\ \hline
      $0-+$ & ${\cal C} \la p_3 | \left( \frac{1}{M_Z^2} \slashed{r}_2 + \frac{M_Z}{2 p_{23}} \slashed{p}_3 \right) (v_f - a_f \gamma^5 ) | p_4 ] = {\cal C} \frac{v_f + a_f}{M_Z} \la r_2 p_3 \ra [ r_2 p_4 ] $\\ \hline
    \hline
  \end{tabular}
  \caption{Helicity Amplitudes for the 3-body Higgs decay  $h\to Zf\bar{f}$.}
  \label{table5}
 \end{center}
\end{table}

Therefore 
\begin{equation}\label{equ:13}
	 \left\la| {\cal M}|^2\right\ra = 8 {\cal C}^2 \left( |v_f|^2 + |a_f|^2 \right) \left[ - p_{34} + \frac{(r_2 \cdot p_3)(r_2 \cdot p_4)}{M_Z^2} \right] ,
\end{equation}
From momentum conservation we can obtain; $2 p_{23} = M_Z^2 - M_H^2 - 2 p_{24} - 2 p_{34}$.
Substituting everything in Eq.(\ref{equ:13}) we obtain 
\begin{equation}
	\left\la \left| {\cal M} \right|^2 \right\ra = \sum \left| {\cal M} \right|^2 = 8 {\cal C}^2 \left( |v_f|^2 + |a_f|^2 \right) \left[ p_{24} - p_{34} - \frac{2 p_{24}}{M_Z^2} \left( p_{24} + p_{34} + \frac{M_H^2}{2} p_{24} \right) \right] .
\end{equation}
Again, from this amplitude we can arrive to the decay width.
\section{Conclusions}\label{conclussions}
In this paper we have presented a summary of the basic formulae of the SHF.  In order to appreciate the value of the methods, we studied the phenomenology of the Electroweak sector of the Standard Model of Particle Physics, including the evaluation of the decays: $Z\to ff$, $h\to ff$, $h\to W^{-}W^{+}$ and $h\to V f'\bar{f}$. 
Although these results are well known, it can be appreciated that the simplification obtained by using SHF, make it worth to use them in teaching of Particle Physics within a modern approach.
\section{Acknowledgements}
The authors wish to thank economic support from CONACYT,  L.D.C. was supported under the grant 220498, B.L. thanks to Fernando Febres Cordero for his valuable comments at the initial stage of this work.

\end{document}